\documentclass[twocolumn
,preprintnumbers,secnumarabic,amsmath,amssymb,address,nofootinbib]{revtex4-2}

\pdfoutput=1
\usepackage{amssymb,amsmath,latexsym,mathrsfs}
\usepackage{url}
\usepackage{enumitem}
\usepackage{bm}
\usepackage{graphicx}
\usepackage{booktabs}
\usepackage[usenames,dvipsnames]{color}
\usepackage[breaklinks,colorlinks,urlcolor=magenta,citecolor=magenta,linkcolor=magenta]{hyperref}
\usepackage{multirow}
\usepackage{float}
\usepackage{cases}
\usepackage{blindtext}
\usepackage{pifont}
\usepackage{hhline}
\usepackage{siunitx}
\usepackage{braket}
\usepackage[table,xcdraw]{xcolor}
\definecolor{linkcolor}{rgb}{0.0, 0.47, 0.75}
\definecolor{citecolor}{rgb}{1.0, 0.5, 0.0}
\definecolor{linkcolor}{rgb}{0.390625,0.5607843137,0.99609375}
\hypersetup{
  linkcolor  = linkcolor,
  citecolor  = linkcolor,
  urlcolor   = linkcolor,
  colorlinks = true
}
\hypersetup{
  linkcolor  = linkcolor,
  citecolor  = linkcolor,
  urlcolor   = linkcolor,
  colorlinks = true
}
\usepackage{verbatim}
\usepackage{hyperref}
\usepackage[normalem]{ulem}
\usepackage{accents}
\usepackage{tikz}
\newcommand\ignore[1]{}			

\usepackage{tocloft}
\usepackage{slashed}
\usepackage{float}
\usepackage{fullpage}
\usepackage{url}

\usepackage{tabularx}

\usepackage{natbib}

\usepackage{upgreek}
\usepackage{outlines}
\def\0{{(0)}}
\def\1{{(1)}}

\usepackage{bm}

\def\ccccend{\end{array}\right)}

\usetikzlibrary{matrix,arrows,decorations.pathmorphing}
\usetikzlibrary{patterns}
\usetikzlibrary{shapes.misc}
\usetikzlibrary{trees}
\usetikzlibrary{decorations.pathmorphing}
\usetikzlibrary{shapes.geometric}
\usetikzlibrary{positioning}
\usetikzlibrary{decorations.markings}
\usetikzlibrary{positioning,arrows}

\usetikzlibrary{decorations.pathreplacing}
\usetikzlibrary{decorations.pathmorphing}
\usetikzlibrary{shapes}
\usetikzlibrary{arrows.meta}
\usetikzlibrary{plotmarks}
\usetikzlibrary{decorations.markings}
\tikzset{
particle/.style={thin,draw=black, postaction={decorate},
decoration={markings,mark=at position .5 with {\arrow[black, line width=0.5mm]{stealth}}}},
gluon/.style={decorate, draw=black, decoration={coil,amplitude=4pt, segment length=5pt}},
photon/.style={decorate, decoration={snake}},
singularity/.style={decorate, draw=black, decoration=zigzag}
}

\usepackage[bottom]{footmisc}

\usepackage{amsthm}
\theoremstyle{theorem}

\begin{document}
	 
\title{A Supernova Constraint on F-theory}

\author{Sebastian Vander Ploeg Fallon$^{a}$}
\author{James Halverson$^{b,c}$}
\author{Liam McAllister$^{a}$}
\author{David J. E. Marsh$^{d}$}

\affiliation{$^{a}$Department of Physics, Cornell University, Ithaca, NY 14853, USA}
\affiliation{$^{b}$Department of Physics, Northeastern University, Boston, MA 02115, USA}
\affiliation{$^{c}$The NSF AI Institute for Artificial Intelligence and Fundamental Interactions}
\affiliation{$^{d}$Department of Physics, King's College London, London, WC2R 2LS, United Kingdom}

\begin{abstract}

We study constraints on the quantum chromodynamics (QCD) axion in F-theory, a strongly coupled limit 
of string theory.
We build models of QCD from compactifications to four dimensions on elliptic fibrations over toric threefolds $B_3$, characterized by an integer $N=h^{1,1}(B_3)$. The QCD axion mass increases  
with $N$, and we find that 
models with sufficiently high $N$ are
inconsistent with observations of the neutrino burst from supernova 1987A, 
disfavoring large regions of the  
moduli space.
Specifically, 
at least $95\%$ of models with $N \ge 8{,}791$ --- a regime that arguably contains the vast majority of known F-theory topologies --- have a QCD axion mass $m_{\text{QCD}}>15~\text{meV}$ and are thus constrained.
This limit is independent of cosmology.
We only consider weakly-curved threefolds, where $\alpha'$ corrections are plausibly negligible.

\end{abstract}

\maketitle

\emph{Introduction:} The quantum chromodynamics (QCD) axion is a compelling candidate for particle physics beyond the Standard Model (SM), providing a solution to the strong CP problem~\cite{Peccei:1977hh,Wilczek:1977pj,Weinberg:1977ma} and a potential component of dark matter~\cite{Dine:1982ah,Preskill:1982cy,Abbott:1982af,Marsh:2024ury}.  
Nearly 50 years after it was first proposed, the theory remains consistent with a wide range of experimental and observational constraints~\cite{Marsh:2015xka,Raffelt:2006cw,Caputo:2024oqc,Carenza:2024ehj,OHare:2024nmr,ParticleDataGroup:2024cfk}, and an experimental program has now grown that has promise to discover it~\cite{Chadha-Day:2021szb,Semertzidis:2021rxs,Adams:2022pbo}. The QCD axion is essentially a one-parameter model~\cite{DiLuzio:2020wdo} defined by the decay constant, $f_a$, which in turn sets the mass~\cite{GrillidiCortona:2015jxo}:
\begin{equation}
    m_a = 5.7\, \left(\frac{10^9 \text{ GeV}}{f_a}\right)\text{ meV}\, .
    \label{eqn:QCD_axion_mass}
\end{equation}

One of the most significant constraints on the properties of the QCD axion is the upper limit on its mass set by the observed duration of the neutrino burst from supernova (SN) 1987A~\cite{Turner:1987by,Mayle:1987as,Burrows:1988ah,Raffelt:1990yz,Lella:2023bfb,Caputo:2024oqc,Carenza:2024ehj}. This limit has been revisited and reassessed for nearly 40 years, consistently implying that $m_a\lesssim 15\text{ meV}$ or equivalently $f_a\gtrsim 4\times 10^8\text{ GeV}$. This is one of the most robust pieces of knowledge about the QCD axion:  
it  stems from  
SM initial states, without assuming axion dark matter,
and it relies on the defining coupling between the axion and gluons.
 
String theory provides
ultraviolet completions that unite general relativity, the SM, and the QCD axion. However, the four-dimensional quantum gravity effective field theories  
obtained from string compactifications generally involve multiple axions, an \emph{axiverse}~\cite{Witten:1984dg,Svrcek:2006yi,Conlon:2006tq,Arvanitaki:2009fg,Acharya:2010zx,Cicoli:2012sz}. 
Recent years have seen significant advances in construction and analysis of axiverses in explicit string compactifications.  This has allowed detailed exploration of the connection between topology and such diverse physics as birefringence, black hole superradiance, cosmic reionization, structure formation, X-ray galaxies, axion-photon couplings, and early Universe cosmology~\cite{Halverson:2019kna,Halverson:2019cmy,superradiance,Demirtas:2021gsq,Broeckel:2021dpz,glimmers,Yin:2025amn,fuzzy,Jain:2025vfh,fallon2025ftheoryaxiverse,Chadha-Day:2021uyt,Bauer:2026nne,deGiorgi:2025ldc,Gendler:2024adn,Agrawal:2025rbr,Reig:2025dqb,Cicoli:2021gss,Leedom:2025mlr}. 

Two observations motivate this work. First,
since the properties of the QCD axion are driven by the decay constant alone,
tests of the QCD axion can be used to make strong  
connections between phenomenology and compactification topology~\cite{glimmers,Gendler:2024adn,Jain:2025vfh}.  
Second, 
in the case of F-theory~\cite{Vafa:1996xn} the number $N$ of axions can be $>10^5$, much larger than is possible in any known solution of weakly-coupled string theory. 
Theories of many axions have distinctive phenomenology, and we will show that structures resulting from ultraviolet completion lead to strong constraints.

In this work we study the QCD axion in F-theory. 
F-theory is described by a Calabi-Yau (CY) fourfold that is an elliptic fibration over a K\"ahler threefold $B_3$. Axions arise in this theory from dimensional reduction of the four-form gauge potential $C_4$ on 4-cycles in $B_3$, giving rise to a number of $C_4$ axions fixed by the Hodge number $h^{1,1}(B_3)$~\cite{Grimm:2010ks}.
We do not consider axions from $C_2$ and $B_2$,
as these fields acquire masses from suitable fluxes, and we neglect potential St\"{u}ckelberg couplings for $C_4$ (cf.~e.g.~\cite{Grimm:2011tb}).

What one gains in the F-theory axiverse~\cite{fallon2025ftheoryaxiverse} compared to the type IIB axiverse~\cite{Demirtas:2018akl,Demirtas:2021gsq,glimmers} is twofold. Firstly, one is no longer restricted to threefolds, immensely widening the available options: as we will discuss, the number of elliptically fibered fourfolds we are able to construct dwarfs the number of known CY threefold constructions~\cite{Yinan}.  Secondly, as is well known, in F-theory some of the four-dimensional gauge theory is fixed by the compactification due to the presence of so-called non-Higgsable clusters (NHC) \cite{Grassi_2015, Morrison_2015, Morrison:2012np}. Relative to type IIB, this provides a new handle for computing  
gauge sectors, including QCD. Thus, in the F-theory axiverse, one widens the lamppost under which one is studying the landscape.

The 
axion effective theories studied in this work are obtained from compactifications of F-theory on elliptic fibrations over certain toric threefold bases $B_3$.   In our constructions, we require  $B_3$ to be weakly curved, in order to avoid effects of unknown $\alpha'$ corrections.  Moreover, we select cases admitting a gauge group containing $\text{SU}(3)\times \text{SU}(2)$, with appropriate couplings, as a proxy for the SM.   In each case, reduction of $C_4$ furnishes a candidate QCD axion. 

\begin{figure}
    \centering
    \includegraphics[width=\linewidth]{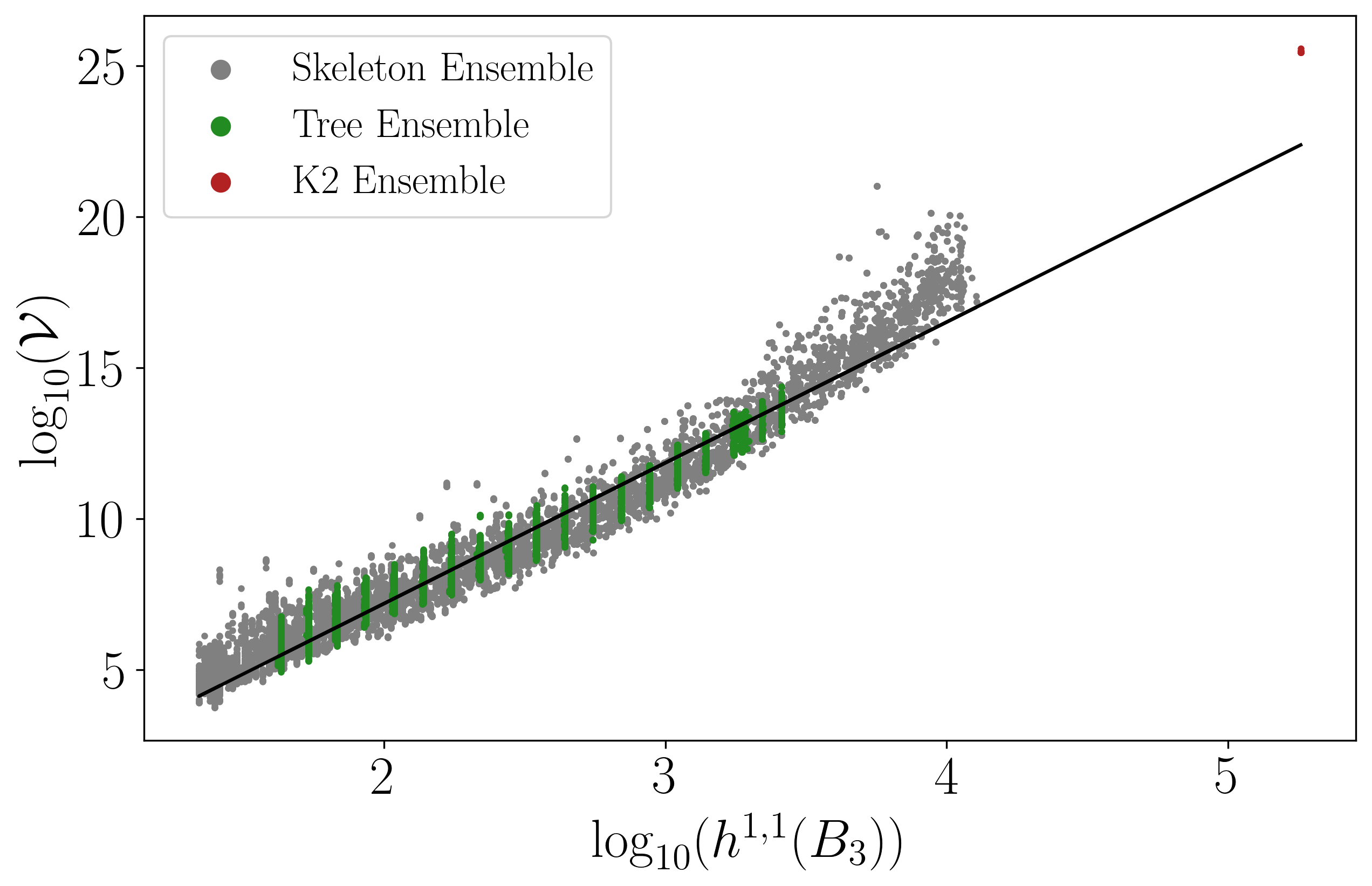}
    \caption{The Einstein-frame volume of the base as a function of $h^{1,1}$. The fit is $y = 4.7 x - 2.1$.}
    \label{fig:overall_vol}
\end{figure}

In this setting we find, along similar lines to \cite{Demirtas:2018akl}, that topologically complex $B_3$ that are weakly curved are necessarily large, with $\mathcal{V}=\mathrm{Vol}(B_3) \gtrsim N^{5}$: see Fig.~\ref{fig:overall_vol}.  Thus, at large $N$ all physical scales in the four-dimensional theory --- including $f_a$ --- become small in Planck units. We show that as a result, for sufficiently large $N$ the SN1987A upper limit on the mass of the QCD axion gives relatively model-independent exclusions of parts of our ensemble.

We do not study moduli stabilization in this work.  Instead, we give evidence that the SN1987A limit excludes  large regions of moduli space (the regions where $B_3$ is weakly curved) in fourfolds with $N \ge 8{,}791$, and hence excludes models realized in such regions, whatever the mechanism of moduli stabilization. While we do not have precise counts of the number of geometries as a function of $N$,  we expect that the vast majority sit above the bound and are thus ruled out. For instance, the Tree ensemble realizes \cite{trees} $\mathcal{O}(10^{755})$ geometries with $N\leq 2{,}591$, but it is estimated \cite{Yinan} that $N=181{,}200$ realizes $10^{45{,}767}$ geometries.

In our ensemble, we find that the contributions to the QCD axion potential from Euclidean D3-branes are subdominant compared to the effects of QCD itself.\footnote{A single model at $h^{1,1}=37$ in our ensemble is dominated by Euclidean D3-branes, and in this case the QCD axion is of poor quality.}  Thus, the QCD axion remains light and hence solves the strong CP problem, 
a situation that has been shown to be generic in weakly-curved CY threefolds at large $N$ \cite{Demirtas:2021gsq}.

\emph{Methodology:}  
The relevant part of the axion effective Lagrangian at the Kaluza-Klein (KK) scale is \cite{fallon2025ftheoryaxiverse}:
\begin{align}
\mathcal{L} & \supset -\frac{1}{2}K_{ij}(\partial_\mu \theta^i)(\partial^\mu \theta^j) - \sum_{i = 1}^N \Lambda_i^4 \cos(2 \pi \theta^i) \nonumber \\ & - \frac{1}{4}G_{\mu\nu} G^{\mu\nu} - \frac{\alpha_{\text{QCD{,}UV}}}{4} Q_{\text{QCD}, i}\theta^i G_{\mu\nu} \tilde{G}^{\mu\nu} + \ldots  \;, \label{eq:axion_eft}
\end{align}
where $i$ runs over prime toric divisors (a basis of 4-cycles) in $B_3$, $K_{ij}$ is the kinetic matrix, $\Lambda_i$ are instanton scales, $\alpha_{\text{QCD{,}UV}}$ is the QCD gauge coupling, $Q_{\text{QCD}, i}$ is the charge vector coupling the axions to QCD, $G_{\mu\nu}$ is the QCD field strength, and we have suppressed the SU(3) trace. Our task is to specify the elements in this Lagrangian based on the compactification data. 

We construct a CY fourfold as a Weierstrass model over a toric threefold $B_3$ given by 
\begin{equation}
y^2 = x^3 + fx z^4 + g z^6\;,
\label{eqn:weierstrass}
\end{equation} 
with $f \in \mathcal{O}_{B_3}(-4K)$ and $g \in \mathcal{O}_{B_3}(-6K)$ (where $\mathcal{O}$ are line bundles and $K$ is the canonical class), and we take $x, y, z$ as coordinates on $\mathbb{P}_{231}$ satisfying $[x : y : z] \sim [\lambda^2 x : \lambda^3 y : \lambda z]$ for $\lambda \in \mathbb{C}^*$. We consider three different Monte Carlo ensembles for generating base manifolds $B_3$: the Tree~\cite{Halverson_2017}, Skeleton~\cite{Taylor_2018} and K2~\cite{Yinan,fallon2025ftheoryaxiverse} ensembles, all of which are discussed in more detail in Ref.~\cite{fallon2025ftheoryaxiverse}. From the toric fan of $B_3$ thus constructed, and using 
Eq.~\eqref{eqn:weierstrass}, one can read off from the singularities the geometrically-non-Higgsable gauge groups $G_{i}$ located on each prime toric divisor $D_{i}$.   

For each geometry, we found all intersecting pairings of prime toric divisors $(D_{i_1}, D_{i_2})$ with
$G_{i_1} \supseteq \text{SU}(3)$ and $G_{i_2} \supseteq \text{SU}(2),$ providing a pseudo-realization of the SM as a product gauge group 
in the UV (we do not consider grand unified theories: for $h^{1,1} > 63$ the string scale is too low for gauge coupling unification, cf.~\cite{Benabou:2026jtv, Leedom:2025mlr}). Flux-breaking and/or a Mordell-Weil $\text{U}(1)$ could potentially realize the full structure. 
For sufficiently large $h^{1,1}$, the pairings we found were overwhelmingly $(G_{i_1}, G_{i_2}) = (\text{G}_2, \text{SU}(2))$.  We next locate a region in K\"ahler moduli space where the $\alpha'$ expansion is plausibly under control and where the divisor volumes give rise to gauge couplings consistent with renormalization group running of the SM. 

Following Ref.~\cite{fallon2025ftheoryaxiverse}, and using Gurobi \cite{gurobi} to solve the associated constraints, 
we first locate a region in K\"ahler moduli space where $B_3$ is weakly curved. Then we fix a choice of divisors hosting QCD and $\text{SU}(2)_L$,
and compute the corresponding  
volumes $\tau_{i_1}$ and $\tau_{i_2}$. 
Associated to each of $\tau_{i_1}$ and $\tau_{i_2}$ is a scale of supersymmetry (SUSY) breaking $M_{\text{SUSY}, i_j}$ such that by using the one-loop beta functions for the SM at energies below $M_{\text{SUSY},i_j}$ and the one-loop beta functions for the minimal supersymmetric SM at energies above $M_{\text{SUSY},i_j}$, one has $\alpha_{G_{i_j}, \text{UV}} = \tau_{i_j}^{-1}$ evaluated at the local KK scale $(m_{\text{KK}})_{i_j} = \frac{g_s}{\sqrt{4 \pi \mathcal{V}}} \frac{2 \pi}{\tau_{i_j}^{1/4}}.$ We adjust the divisor volumes using the constrained optimization techniques of Ref.~\cite{fallon2025ftheoryaxiverse} to solve for a K\"ahler form with  
a consistent global value of $M_{\rm SUSY}$ with $1 \text{ TeV} \leq M_{\text{SUSY}} \leq \min(m_{\text{KK}, i_1}, m_{\text{KK}, i_2})$, and use this point in K\"ahler moduli space to calculate axion observables.\footnote{At $h^{1,1} = 181{,}200$, constrained optimization is
very costly, 
so we instead use a combination of dilation of the point in K\"ahler moduli space and gradient ascent.}  
The number of models thus constructed is described in Table~\ref{table:ensemble}.

From the data of the gauge groups, we specify the instanton scales $\Lambda_i$: 
\begin{equation}
\Lambda_i^4 = \begin{cases}(75.5 \text{ MeV})^4, & i = \text{QCD} \\ \frac{8 \pi W_0 \mathcal{A}_i \tau^i}{\mathcal{V}^2 C_2(G_i)} e^{-2 \pi \tau^i/C_2(G_i)} e^{K_\text{other}} , & i \neq \text{QCD}\end{cases}\;,
\end{equation} 
where $C_2(G_i)$ is the dual Coxeter number of $G_i$, which we set to $1$ in the case that $D_i$ lacks a non-Higgsable gauge group, $W_0$ is the Gukov-Vafa-Witten superpotential, $\mathcal{A}_i$ is the appropriate Pfaffian prefactor, $\mathcal{V}$ is the overall volume of $B_3$, and we take $\mathrm{exp}(K_\text{other}) = \frac{g_s^4}{128}$ with $g_s$ the string coupling~\cite{fuzzy}. We set $g_s = W_0/M_{\text{pl}}^3 = \mathcal{A}_i/M_{\text{pl}} = 1$ for all $i$: this does not strongly affect our conclusions, which are driven by $K_{ij}$ and the QCD instanton scale.

In general, computing the masses and couplings of all the axions is a numerically difficult task.  
For the purposes of the present work, we only require an accurate estimate of the QCD axion mass, which we can find using a modified perturbative approach following Refs.~\cite{Demirtas:2021gsq,glimmers,fallon2025ftheoryaxiverse}, in which we only retain a subset of the $\Lambda_i$ in a block near 75 MeV.

\emph{SN1987A and the QCD axion:} The core-collapse of SN1987A is believed to have led to the formation of a proto-neutron star (PNS), and recent JWST observations support this~\cite{Fransson:2024csf}. This PNS cooled over time, emitting neutrinos, which were observed by several detectors around the world over the course of 10 seconds in 1987~\cite{PhysRevLett.58.1494,PhysRevLett.58.1490,Alekseev:1988gp}. Axions interact more feebly than neutrinos, and if they were produced during the supernova then they would cause more rapid cooling of the PNS, and a consequently shorter neutrino burst.
Thus, the measured burst excludes axions with large couplings to nucleons~\cite{Turner:1987by,Mayle:1987as,Burrows:1988ah} (for reviews, see Refs.~\cite{Raffelt:2006cw,Caputo:2024oqc,Carenza:2024ehj}). 
We consider axion emission via nuclear bremsstrahlung, $N+N\rightarrow N+N+a$, mediated by the axion-nucleon coupling. Roughly, axions are considered to be excluded if the emissivity in axions, $\varepsilon_a$, is greater than the neutrino emissivity, $\varepsilon_\nu\approx 10^{19}\text{ erg g}^{-1}\text{s}^{-1}$, at which point the neutrino burst duration is approximately halved. Taking $\varepsilon_a<\varepsilon_\nu$ to set the limit is known as the ``Raffelt criterion'' following Ref.~\cite{Raffelt:1990yz}, and is based on a series of one-dimensional supernova simulations.

\begin{table}
\begin{center}
\begin{tabular}{ |c|ccc| } 
 \hline
 Ensemble  & Tree & Skeleton & K2 \\
 \hline 
 Min. $h^{1,1}$  & 35 & 10 & 2{,}560\\ 
 Max. $h^{1,1}$  & 2{,}591 & 15{,}415 & 181{,}200 \\ 
 Total Models  & 14{,}627 & 16{,}556 & 54 \\ 
 \hline
\end{tabular}

\caption{Ensemble data. 
Number of models corresponds to number of gauge group choices for QCD, with gauge couplings fixed in the stretched K\"ahler cone as described in the text.}\label{table:ensemble}

\end{center}

\end{table}

For a single nucleon species the axion emissivity can be estimated as~\cite{Raffelt:2006cw} $\varepsilon_a\approx C_{aNN}^2(m_a/f_a)^2 (T^4/4\pi^2 m_N)^3$, with $T$ the supernova temperature and $m_N$ the nucleon mass. The dimensionless axion-nucleon coupling is defined by $\mathcal{L}_{\rm int} = \frac{C_{aNN}}{2f_a}(\partial_\mu a)\bar{N}\gamma^\mu\gamma_5 N$. The axion-gluon coupling alone generates $C_{app}=-0.47(3)$, $C_{ann}=-0.02(2)$~\cite{GrillidiCortona:2015jxo}. Since $C_{ann}$ is consistent with zero within errors, we conservatively set it to zero. Using these values, and following Refs.~\cite{Carenza:2019pxu,Caputo:2024oqc} for a more accurate estimate of the emissivity, the Raffelt criterion can be re-arranged to constrain the axion mass:  
\begin{equation}
    m_a<15\text{ meV} \Leftrightarrow f_a>4\times 10^8\text{ GeV}\, .
    \label{eqn:axion_mass_bound}
\end{equation}
This approximate limit is close to many estimations made over the decades, and in particular to the recent work of Ref.~\cite{Lella:2023bfb}, which finds $m_a\lesssim 10\text{ meV}$ after including in addition pion-nucleon scattering.\footnote{All neutron stars cool by axion emission, and observations of the ``Magnificent Seven'' place a similar bound on the axion mass to that from SN1987A~\cite{Buschmann:2021juv}.}  
A QCD axion mass exceeding these limits by any significant amount is clearly in conflict with the observed and simulated astrophysics of SN1987A, and such models can be confidently excluded.

\begin{figure}
    \centering
    \includegraphics[width=0.95\linewidth]{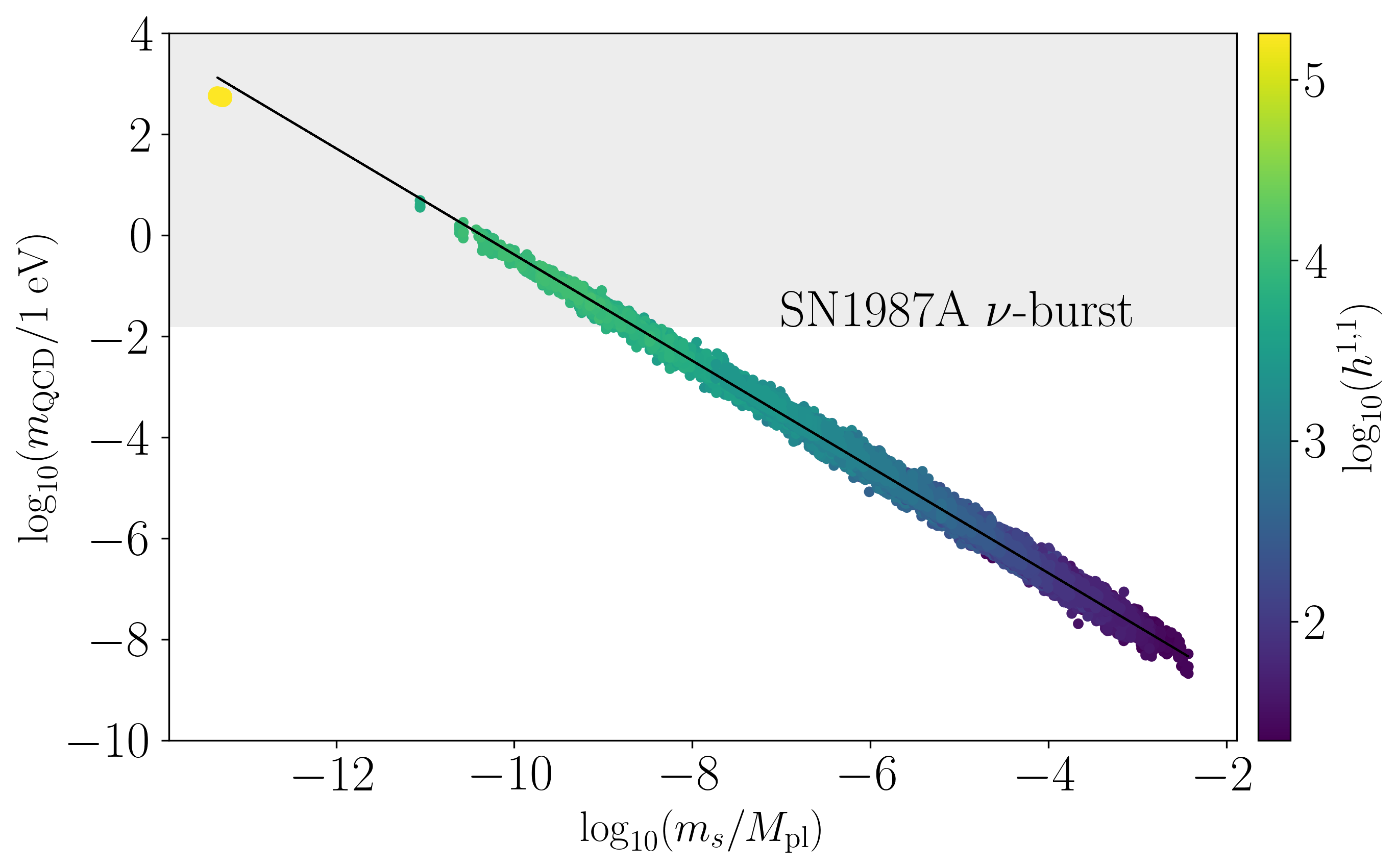}

\caption{The mass of the QCD axion as a function of the string scale $m_s$. Data in the shaded region are in tension with the neutrino burst from SN1987A.}
\label{fig:mass_ms}
\end{figure}

\begin{figure}
    \centering
    \includegraphics[width=0.95\linewidth]{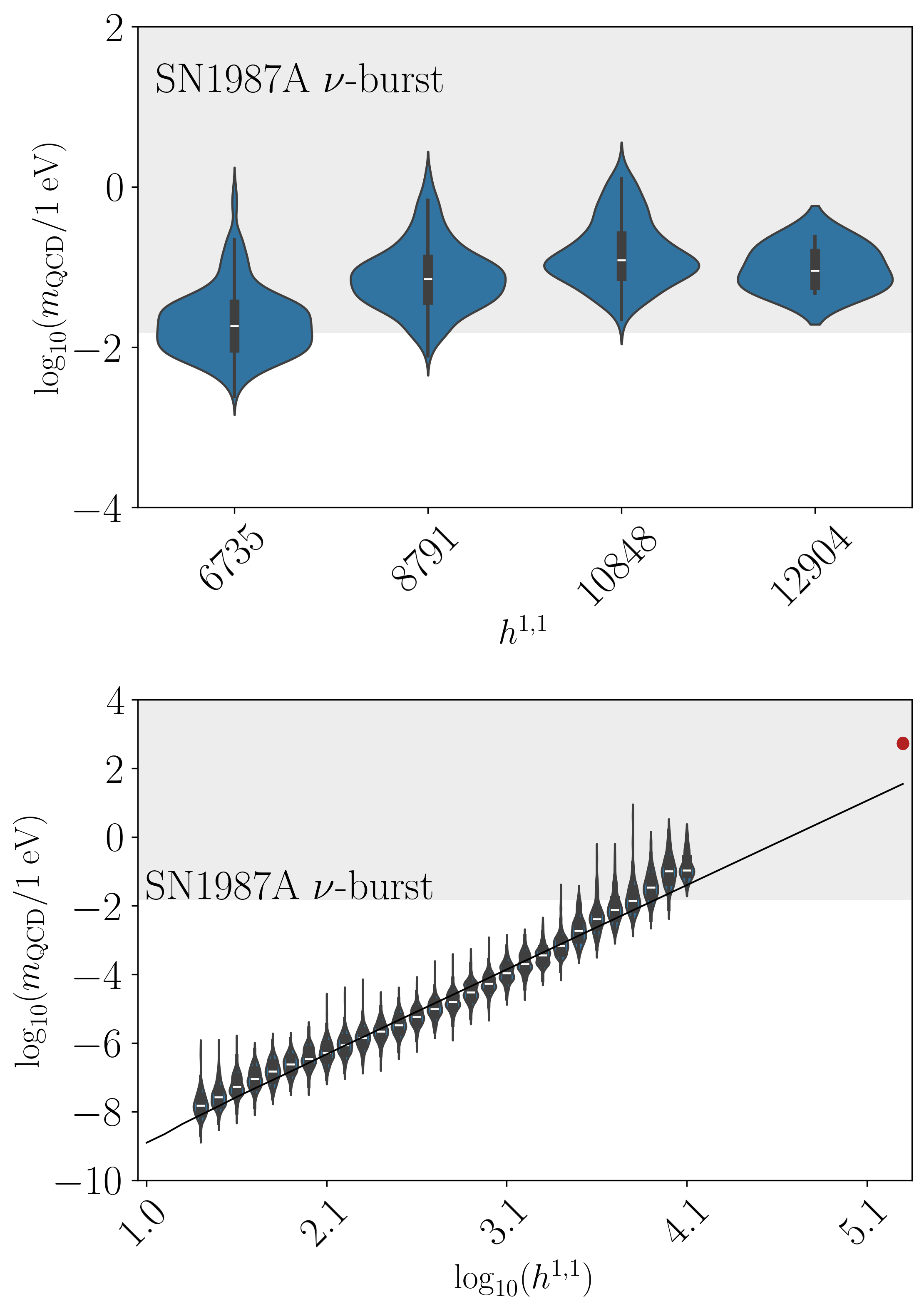}

\caption{The mass of the QCD axion as a function of the number of axions, $h^{1,1}$. Data in the shaded region are in tension with the neutrino burst from SN1987A. \textit{Top:} Linear bins around the exclusion. \textit{Bottom:} All data binned with 0.1 dex. Red dot marks the data at $h^{1,1}=181{,}200$.}
\label{fig:mass_h11}
\end{figure}

\emph{Results:}   
To set constraints using \eqref{eqn:axion_mass_bound},
we computed the distribution of QCD axion masses in our ensemble.
Letting $m_s = \frac{g_s}{\sqrt{4 \pi \mathcal{V}}} M_{\rm pl}$
be the string scale, we find \begin{equation}\log_{10}(m_{\text{QCD}}/1 \; \text{eV}) = -1.05 \log_{10}(m_{\text{s}} / M_{\text{pl}}) - 10.89\;, \label{eq:m_qcd_m_string}\end{equation} 
and \begin{equation}\log_{10}(m_{\text{QCD}}/1 \; \text{eV}) = 2.46 \log_{10}(h^{1,1}) - 11.45 \;.\end{equation} 
The slope in equation \eqref{eq:m_qcd_m_string} agrees very well with \cite{Cheng:2025ggf}, which predicts a slope of $-1$.

Our main result can be seen in Fig.~\ref{fig:mass_h11}: every $h^{1,1}$ bin with $h^{1,1} \geq 8{,}791$ has $m_{\text{QCD}} > 15 \text{ meV}$ for at least $95\%$ of models.  In this sense, models with $h^{1,1} \geq 8{,}791$ are in tension with the SN1987A limit \eqref{eqn:axion_mass_bound}.

\emph{Discussion:} We constructed a large class of F-theory compactifications on Weierstrass models over toric threefolds $B_3$, incorporating non-Higgsable gauge groups furnishing pseudo-realizations of the Standard Model, as well as a QCD axion from $C_4$.  We studied the regions in K\"ahler moduli space where $B_3$ is weakly-curved, so that the $\alpha'$ expansion is plausibly controlled.  In these regions, the mean QCD axion decay constant $f_a$ decreases with $N=h^{1,1}(B_3)$, and the SN1987A upper limit on the QCD axion mass provides a stringent constraint on models with large $N$ (see Fig.~\ref{fig:mass_h11}), which we recall are expected to represent the vast majority of F-theory geometries~\cite{Yinan}.

We did not study moduli stabilization, and instead identified regions of moduli space that are in tension with SN1987A.
However, we have in mind the possibility of stabilization by perturbative corrections to the K\"ahler potential. Stabilization by nonperturbative superpotential terms, as in \cite{Kachru:2003aw}, is well-known (e.g.~Ref.~\cite{Conlon:2006tq}) to be in tension with the QCD axion solution of the strong CP problem, and moreover selects regions in moduli space where the volume grows even more rapidly with $h^{1,1}$ than in the regions we study. At large $h^{1,1}$, such nonperturbatively-stabilized models likely face further tensions with experimental constraints, e.g. from the scale of supersymmetry breaking and the scale of the moduli potential. 

Even perturbatively stabilized models face constraints. Specifically,
the large-$N$ models studied here have large volumes $\mathcal{V}$, and correspondingly low Kaluza-Klein scales.  
As a result, the scale of the moduli potential is also small at large $N$, and thus so are the saxion masses.
Indeed, for $N \gtrsim 10{,}000$, the saxion parameterizing the overall volume has $m \sim 40 \text{ GeV}$, 
and so will decay after Big Bang nucleosynthesis.
Thus, some of the large-$N$ models that are excluded by SN1987A may \emph{also} be ruled out, but in a model-dependent way, by the cosmological moduli problem.

We only considered  supernova emission of the QCD axion, via the gluon coupling. In reality, any other axions coupled to gluons, and with
$m_a<T_{\rm SN}\approx 30\text{ MeV}\,,$ 
would also be emitted from SN1987A by nuclear bremsstrahlung and would contribute to the axion emissivity and further shorten the neutrino burst. This would strengthen the constraints, making our limits ignoring these heavier axions conservative.
Moreover, other axions in the spectrum could cause a host of further problems, such as overproduction of dark matter.
Although these 
further constraints could be strong, their impact depends on  cosmological assumptions, and so we defer such issues to future work.

Our visible-sector model-building was incomplete: we arranged for a gauge group containing the nonabelian factors of the SM, but did not ensure a suitable $\text{U}(1)$ factor, nor did we consider supersymmetry breaking.\footnote{The couplings of the QCD axion in examples of F-theory models containing the SM, with small values of $h^{1,1}$, were studied in \cite{Chen:2026bxp}. See also  \cite{Nee:2026msa} for a recent study of axion-photon couplings in F-theory models with grand unified gauge groups.}
Furthermore,
we have not modeled axion couplings to fermions, although these can in general be present.
Direct couplings between the axion and the SM quarks change the values of $C_{ann}$ and $C_{app}$ relative to those induced by gluons alone, with the down (up) quark coupling appearing with opposite sign in the neutron (proton) coupling compared to the gluon-induced coupling and those of all the other quarks~\cite{GrillidiCortona:2015jxo}. A tuning of the axion-quark couplings (which we are omitting) with $C_{auu}+C_{add}\approx 1$ leads to an approximate cancellation and a relaxation of the bound given by Eq.~\eqref{eqn:axion_mass_bound}. It would be interesting to try to construct such couplings in F-theory that evade our limit.
Finally, it is possible that a realization of QCD 
on a cycle
with vanishing self-intersection can allow for a region of moduli space in which $f_a$ remains large even as $\mathcal{V}$ increases: cf.~\cite{Cicoli:2012sz,fuzzy}.
Such a region of moduli space would evade our constraints. We have not seen such examples realized in random sampling, and leave an exploration of this potential mechanism for future work.

\emph{Conclusions:}
New computational methods~\cite{fallon2025ftheoryaxiverse} have allowed us to make contact between F-theory and the effect the QCD axion has on neutrino emission from supernovae. 
We have shown that complex F-theory topologies --- which dominate our statistical ensemble --- yield large QCD axion masses in the part of their K\"ahler moduli space where the internal space is weakly curved.  For $h^{1,1} \ge 8{,}791$, 
$\ge 95\%$ of models are in 
conflict with observations of neutrinos from SN1987A. 

This work motivates further study of the detailed properties of the QCD axion in F-theory.

\acknowledgements

We are grateful to Andrea Caputo, Pang Yen Chen,
Naomi Gendler, Jakob Moritz, Mario Reig, Ben Safdi, and Elijah Sheridan for helpful discussions.  Software development was assisted by Anthropic's Claude Opus 4.7.  
D.J.E.M. is supported by an Ernest Rutherford Fellowship (Grant No.~ST/T004037/1) and a consolidator grant (Grant No.~ST/X000753/1) from the Science and Technologies Facilities Council, United Kingdom.
The research of L.M. and S.V.P.F. is supported in part by NSF grant PHY-2309456. J.H. is supported by NSF CAREER grant
PHY-1848089 and PHY-2209903. This work was completed in part using the Explorer Cluster, supported by Northeastern University's Research Computing team.  J.H. and L.M. acknowledge the Aspen Center for Physics, which is supported by National Science Foundation grant PHY-2210452, for providing a productive environment for the completion of this work. 

\bibliographystyle{utphys}
\bibliography{refs}

\end{document}